\documentclass[9pt]{vgtc}                           % final (conference style)
\ifpdf%                                % if we use pdflatex
  \pdfoutput=1\relax                   % create PDFs from pdfLaTeX
  \usepackage{graphicx}                % allow us to embed graphics files
  \DeclareGraphicsExtensions{.pdf,.png,.jpg,.jpeg} % for pdflatex we expect .pdf, .png, or .jpg files
\else%                                 % else we use pure latex
  \usepackage{graphicx}                % allow us to embed graphics files
  \DeclareGraphicsExtensions{.eps}     % for pure latex we expect eps files
\fi%

\graphicspath{{figures/}{pictures/}{images/}{./}} % where to search for the images

\usepackage{microtype}                 % use micro-typography (slightly more compact, better to read)
\PassOptionsToPackage{warn}{textcomp}  % to address font issues with \textrightarrow
\usepackage{textcomp}                  % use better special symbols
\usepackage{mathptmx}
\usepackage{amsmath}
\usepackage{times}                     % we use Times as the main font
\usepackage{cite}                      % needed to automatically sort the references
\usepackage{tabu}                      % only used for the table example
\usepackage{booktabs}   
\usepackage{graphicx}
\usepackage{comment}
\usepackage[subtle]{savetrees} 
\usepackage{subcaption}
\usepackage{xcolor}
\usepackage{lipsum} % for dummy text% only used for the table example
\onlineid{0}

\vgtccategory{Research}

\title{Optical Tag-Based Neuronavigation and Augmentation System for Non-Invasive Brain Stimulation}

\author{
    Xuyi Hu\thanks{e-mail: xh365@cam.ac.uk} \hspace{1.5cm}
    Ke Ma\thanks{e-mail: km834@cam.ac.uk} \hspace{1.5cm}
    Siwei Liu\thanks{e-mail: sl2049@cam.ac.uk} \hspace{1.5cm}
    Per Ola Kristensson\thanks{e-mail: pok21@cam.ac.uk} \hspace{1.5cm}
    Stefan Goetz\thanks{e-mail: smg84@cam.ac.uk} \\
    \vspace{-3pt} \\ % Adds extra space
    \scriptsize University of Cambridge
}

\let\oldtwocolumn\twocolumn
\renewcommand\twocolumn[1][]{%
    \oldtwocolumn[{#1}{
    \begin{center}
           \includegraphics[width=0.85\textwidth]{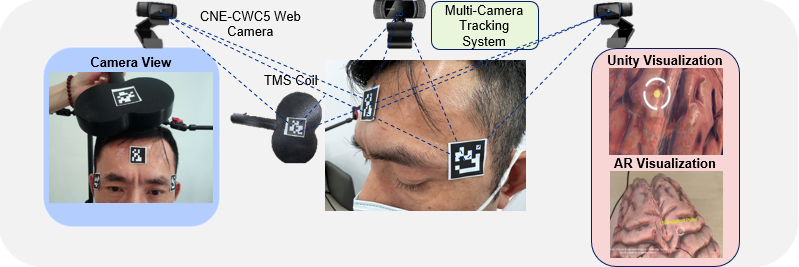}
           \captionof{figure}{The optical tag-based neuronavigation system combines multi-camera tracking with augmented reality (AR) visualization to improve transcranial magnetic stimulation (TMS) procedures. This system uses low-cost web cameras to track black-and-white optical fiducials on the TMS coil and patient for real-time tracking. Unity 3D visualization and AR overlays guide precise coil positioning and target identification.}
           \label{fig:flowchart}
           \vspace{-10pt}
        \end{center}
    }]
}

\abstract{Accurate neuronavigation is critical for effective transcranial magnetic stimulation (TMS), as stimulation outcomes depend directly on precise coil placement. Existing neuronavigation systems are often costly, complex, and prone to tracking errors. To address these limitations, we present a computer-vision-based neuronavigation system that enables real-time tracking of the patient and TMS instrumentation. The system integrates a multi-camera optical tracking setup with consumer-grade hardware and visible markers to drive a digital twin of the stimulation process. A dynamic 3D brain model in Unity updates in real time to visualize coil position and estimated stimulation targets. Augmented reality (AR) is further incorporated to project this model directly onto the patient’s head, enabling intuitive, in-situ coil adjustment without reliance on abstract numerical displays. Overall, the proposed approach improves spatial precision and accuracy while enhancing usability.
} % end of abstract

\CCScatlist{ 
 Augmented Reality, Computer Vision, Visualization, Neuronavigation, Transcranial Magnetic Stimulation}

\begin{document}
\firstsection{Introduction}
\maketitle
An exciting area of augmented reality (AR) is to better support therapy and research on the human brain. 
Transcranial magnetic stimulation (TMS) is a focal non-invasive technique to write signals into circuits of the brain through strong brief electromagnetic pulses across the intact scalp and skull \cite{barker1985non,wassermann2008oxford,liu2025prior}. In contrast to deep brain stimulation \cite{perlmutter2006deep,ma2024extraction}, intracortical microstimulation \cite{flesher2016intracortical,ma2023correlating}, or epidural stimulation \cite{edgerton2011epidural}, TMS does not require any surgical implantation or even anesthesia. Instead, it works on awake patients.  A strong current pulse in a focal stimulation coil generates a local magnetic field~\cite{deng2017,ma2025optimal}. This magnetic field penetrates the scalp as well as skull with minimal distortion and induces a spatially confined electric field in the brain, which in turn depolarizes neurons to respond with so-called action potentials, that is, a neural signal.

TMS is widely used in the treatment of neurological and psychiatric disorders, including depression~\cite{sonmez2019accelerated,perera2016clinical, zhou2024revisiting}, obsessive-compulsive disorder (OCD)~\cite{rodriguez1996transcranial,ma2023exploring}, and chronic pain~\cite{galhardoni2015repetitive,hamid2019noninvasive,wang2025stochastic}. Cortical mapping and pre-surgical planning locates functions in the brain by sending test pulses and generates maps~\cite{lefaucheur2016value,krieg2017protocol}. 

% In depression treatment, stimulation of the left dorsolateral prefrontal cortex (DLPFC) aims to up-regulate the mood-control network~\cite{pascual1996rapid,jahanshahi1998left}. For OCD, TMS targets hyperactive circuits in the cortico-striato-thalamo-cortical (CSTC) loop, reducing compulsive behaviors~\cite{li2016cortico}. In chronic pain management, stimulation of the motor cortex alters pain perception pathways, providing relief to patients with conditions such as neuropathic pain~\cite{lefaucheur2006use,leo2007repetitive} and fibromyalgia~\cite{marlow2013efficacy}. Beyond clinical applications, TMS is also used in neuroscience research to study brain function~\cite{hallett2000transcranial,ma2023correlating} and cognitive processes~\cite{luber2014enhancement,ma2022revised}.

% Neuronavigation is essential in TMS treatment for achieving precise coil positioning, which directly affects stimulation accuracy and clinical efficacy. Misalignment of the TMS coil can lead to inconsistent results, reduced therapeutic effects, and unintended neural stimulation in adjacent circuits. Given the variability in individual brain anatomy, precise targeting is essential for optimizing treatment efficacy and reproducibility~\cite{herwig2001navigation}. Without accurate guidance, TMS  fails to stimulate the intended cortical regions and the  neuromodulation effects are not reliable ~\cite{ruohonen2010navigated}.

Traditional TMS guidance relies on neuronavigation systems based on optical or electromagnetic tracking~\cite{goetz2021oxford}. Optical systems typically use infra-red stereo cameras to track retroreflective markers, requiring pre-calibration, a restricted operating volume, and uninterrupted line of sight~\cite{gumprecht1999brainlab,ganslandt2002neuronavigation}. Guidance is usually presented on external screens, forcing users to interpret numerical data or 3D visualizations from a perspective different from their own, which increases cognitive load and training requirements. 

Augmented reality (AR) provides clinicians with intuitive, real-time spatial guidance by overlaying anatomical models or stimulation targets directly onto the patient’s head, enabling direct alignment of the TMS coil with cortical targets~\cite{frantz2018augmenting,schneider2021augmented}. This immersive feedback reduces reliance on abstract numerical displays and improves positioning accuracy, usability, and reproducibility. AR has been applied to cortical mapping, pre-surgical planning, and psychiatric neuronavigation, demonstrating its potential to enhance clinical workflows and treatment consistency~\cite{wang2023three}.

Computer vision (CV) promises a non-invasive and cost-effective solution for guiding TMS coil placement. In contrast to the heavy retro-reflective balls, trackers can be simple black-and-white patterns cost-effectively printed onto multiple stickers to be visible from many perspectives. Consumer-grade cameras have reached high resolutions and are readily available to allow multiple cameras from many view points to avoid shadow and perspective problems. Further, multiple cameras can compensate for abberation errors.

\begin{figure*}[t]
  \centering
  \begin{subfigure}[b]{0.4\textwidth}
    \centering
    \includegraphics[height=0.6\textwidth]{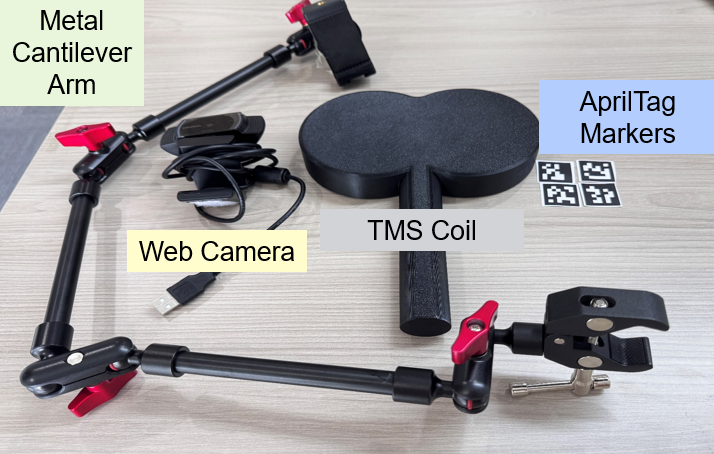}
    \caption{Tools and targets}
    \label{fig:pos1}
  \end{subfigure}
  % \hfill
  \begin{subfigure}[b]{0.4\textwidth}
    \centering
    \includegraphics[height=0.6\textwidth, width=0.9\textwidth]{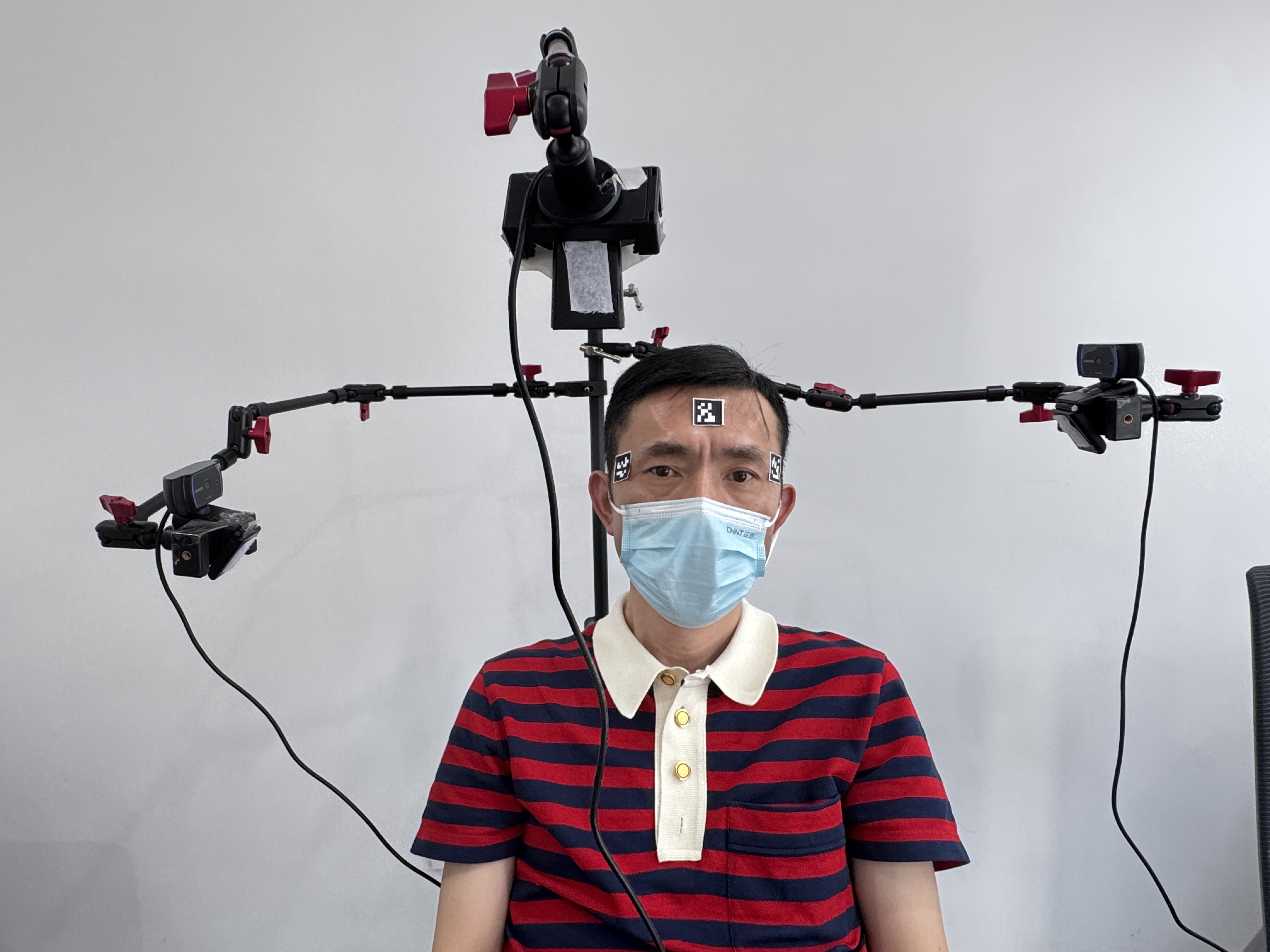}
    \caption{Multi-Camera TMS navigation workstation}
    \label{fig:pos5}
  \end{subfigure}
  \caption{(a) Components and (b) multi-camera TMS neuronavigation system during experiment. The system consists of a TMS coil, a multi-camera tracking setup for real-time spatial localization, and optical markers attached to the tracked objects and the head for position and orientation detection. The cameras provide synchronized tracking data and translate all components into a unified coordinate frame.}
  \label{fig:pipeline}
\end{figure*}

In this paper, we present an AR-enabled neuronavigation system based on printed black-and-white tags, specifically AprilTags, integrated with AR for visualization to enhance accessibility and reduce cost of neuronavigation system (Figure~\ref{fig:flowchart}). This system leverages CV-based tracking to provide real-time, precise spatial localization of the TMS coil and the patient’s head. The primary contributions of this work are the following:
\begin{enumerate}
    \item We present a novel low-cost and easy to use tag-based multi-camera neuronavigation system for real-time tracking of TMS coil positioning within 5 mm accuracy at a total cost of £60.  
    \item We show that the system’s tracking accuracy and effectiveness means this system is practical and competitive compared to existing more expensive neuronavigation methods.
    \item We provide integration of a 3D brain model with Unity-based AR visualization, which offers an interactive and easy to use interface for real-time neuronavigation guidance.
\end{enumerate}

% What is TMS? -> What is the application? -> What is the limitation of current tech? why? (neuronavigation, stimulation accuracy, require well-trained medical expert) -> the current study goal (augmented reality, low-cost, easy-to-use/implementation). 

%%neuron nevigation
\section{Conventional Neuronavigation Systems}

Commercial neuronavigation systems, such as optical and electromagnetic systems, can achieve sub-millimeter accuracy~\cite{de2024precision,yu2024clinical}, but require substantial financial investment, ranging from \$30{,}000 to \$100{,}000~\cite{sulangi2024neuronavigation,asfaw2024charting}. This results in a high accuracy–cost ratio and limits accessibility. In addition, these systems are complex to operate, requiring extensive training and workflow integration. By contrast, recent advances in consumer-grade cameras enable high-resolution, high-frame-rate visual tracking at significantly lower cost~\cite{wang2019pose,choi2018robust}.

Previous work has explored TMS coil guidance using marker-based tracking, robotic systems, or mixed-reality visualization. Washabaugh et al.~\cite{washabaugh2016low} proposed a marker-based approach with acceptable accuracy but limited intuitive feedback and reliance on costly reflective markers. Lin et al.~\cite{lin2019trajectory} introduced a robotic system for coil alignment, though its expensive hardware and complex setup limit portability. Leuze et al.~\cite{leuze2018mixed} focused on mixed-reality visualization for intuitive guidance but did not address stimulation target localization. In contrast, our approach avoids specialized tracking hardware or robotics, providing a lightweight, operator-guided solution with real-time visual augmentation for accurate coil positioning.

\begin{comment}
\subsection{Computer-Vision for Neuronavigation Systems}
%So far, CV-based neuronavigation systems could use marker-based tracking, infrared cameras, and AI-driven pose estimation to improve real-time localization for TMS and neurosurgical applications\cite{robertson2021frameless,chiurillo2023high}.
Head-tracking techniques, depth cameras, and LiDAR-based localization are used to enhance precision while reducing reliance on costly MRI-based navigation. CV based systems  provide a non-invasive, real-time tracking alternative but face challenges related to occlusion, lighting conditions, and calibration errors\cite{strickland2013neuronavigation}.  AR and Mixed Reality (MR) based neuronavigation systems further improve accuracy by providing interactive 3D overlays of brain structures that align with the patient's anatomy in real-time\cite{strickland2013neuronavigation,yavas2021three}. These systems, which leverage HoloLens, Magic Leap, or Virtual Reality (VR) headsets, enable real-time visualization of TMS targets and enhance surgical precision by superimposing anatomical landmarks onto the real-world environment \cite{li2022holographic}. AR assisted neuronavigation allows for more intuitive coil placement and reduces operator dependency, while VR based systems provide immersive preoperative planning environments for optimizing stimulation targets \cite{gouveia2022mixed}. 
\end{comment}

\section{Tag-Based Multi-Camera Tracking System}
\subsection{Tracking Algorithm}
Our system uses regular rectangular black-and-white fiducial patterns from the AprilTag family~\cite{olson2011apriltag,wang2016apriltag}. It consists of high-contrast square patterns that encode unique identification numbers. Further, the sharp edges and the fundamental rectangular grid enable accurate estimation of location and orientation. The tracking algorithm extracts these markers from camera images, decodes their identities, and determines their 3D position and orientation relative to the camera through perspective transformation. 

Tracking based on optical tags with patterns in the visible spectrum provides several advantages over alternative tracking methods. It is computationally efficient and can operate without purpose-designed hardware and instead exploit the latest high-quality low-cost consumer-grade multi-camera and monocular-camera systems. 

% \begin{figure}[!tb]
% \centering
% \includegraphics[width=0.8\linewidth]{pictures/AprilTag.png}
% \caption{\label{fig:AprilTag}Representation of the planar black-and-white tag family (AprilTag tag36h11). The tag's four corner points are labeled for accurate pose estimation. A coordinate frame at the tag center indicates its orientation and position in space, which can be derived from the corner positions even from a single camera perspective.}
% \end{figure}

\subsection{Multi-Camera Setup}
To enable real-time 360° tracking of the TMS coil and patient’s head, we employ a three-camera system that provides continuous spatial localization while minimizing occlusions. Each camera was individually calibrated offline to estimate intrinsic parameters using standard camera calibration procedures. This configuration enables accurate neuronavigation without relying on expensive infrared motion-capture hardware. A shared world coordinate frame is established by combining the fixed geometry of the calibrated cameras with the known spatial arrangement of the head-mounted AprilTag markers. This implicit world anchor enables consistent multi-camera pose estimation without requiring external tracking infrastructure. Each camera captures synchronized frames and independently detects AprilTag markers. Positional information from all views is fused using computer vision to estimate the six-degrees-of-freedom (6~DoF) pose of both the coil and the head. The multi-camera design improves robustness, allowing reliable tracking even when individual cameras experience temporary occlusion. It supports real-time compensation for head motion, making it suitable for practical clinical and research scenarios involving natural patient movement during TMS sessions.

% \begin{figure}[!tb]
% \centering
% \includegraphics[width=0.8\linewidth]{pictures/camera setup.png}
% \caption{\label{fig:camera setup}Multi-Camera system setup.}
% \end{figure}

\subsection{Optical Target Estimation}
Accurate target estimation is critical for precise TMS coil placement. We employ a multi-camera tracking system using AprilTag fiducial markers to estimate the poses of the patient’s head and the TMS coil. Five markers are used, with four attached to the head and one to the coil. A global coordinate frame is established by combining the fixed camera geometry with the known spatial arrangement of head-mounted markers, enabling continuous pose estimation under natural head motion and operator hand movement. The multi-camera and multi-marker design provides redundancy, ensuring stable and robust target estimation even under partial occlusion.

\subsubsection{Direct Linear Transform}
In pose estimation, we aim to determine the rotation and translation that align a known set of 3D points to their corresponding 2D projections. The transformation from 3D camera coordinates to 2D image coordinates follows the camera projection equation
\begin{equation}
\begin{bmatrix} x \\ y \\ 1 \end{bmatrix} = s
\begin{bmatrix} f_x & 0 & c_x \\ 0 & f_y & c_y \\ 0 & 0 & 1 \end{bmatrix}
\begin{bmatrix} X \\ Y \\ Z \end{bmatrix}
=s\mathbf{K}\begin{bmatrix} X \\ Y \\ Z \end{bmatrix},
\label{eq:world_to_camera}
\end{equation}
where \( f_x, f_y \) are the focal lengths in the \( x \) and \( y \) directions, \( c_x, c_y \) define the optical centre of the camera, \( s \) is depth scale factor and \(\mathbf{K}\) is the intrinsic camera matrix.

\subsubsection{Pose Estimation}

% Given a set of known 3D points in the tag’s coordinate frame and their corresponding 2D projections in the image plane, we can write
% \begin{equation}
% \begin{bmatrix}
% x_i \\[4pt]
% y_i \\[4pt]
% 1
% \end{bmatrix}
% =
% \mathbf{K} \Bigl(
%    \mathbf{R}
%    \begin{bmatrix}
%    X_i \\[3pt]
%    Y_i \\[3pt]
%    Z_i
%    \end{bmatrix}
%    \;+\;
%    \mathbf{t}
% \Bigr),
% \end{equation}
% where $\mathbf{R}$ is the rotation matrix, and $\mathbf{t}$ is the translation vector.

Once a fiducial marker is detected, the  algorithm provides the 2D image coordinates of the tag corners. Then, the camera projection matrix as defined in Equation~\ref{eq:world_to_camera}  establishes 3D-to-2D correspondences. Given a set of known 3D landmarks (e.g., AprilTag marker corners) and their corresponding 2D image projections, we estimate the camera pose $\mathbf{R}$ and $\mathbf{t}$ by solving 
\begin{equation}
\min_{\mathbf{R}, \mathbf{t}} \sum_{i} \left\|\begin{bmatrix} x_i \\ y_i \\ 1 \end{bmatrix} - \pi \left(\mathbf{K} \biggl( \mathbf{R} \begin{bmatrix} X_i \\ Y_i \\ Z_i \end{bmatrix} + \mathbf{t}  \biggr)\right)\right\|^2\!\!{},
\label{eq:pose estimation}
\end{equation}
where $\pi(\cdot)$ is the camera projection function.

\subsection{Gaussian Error Combination for Increased Precision}
% For multiple camera views, the final target estimate is obtained by fusing multiple camera distance estimates using weighted averaging:

% \begin{equation}
% \hat{d} = \frac{\sum_{i} w_i d_i}{\sum_{i} w_i}, \quad w_i = \frac{1}{\sigma_i^2}
% \end{equation}

% This method prioritizes measurements with lower uncertainty, improving the accuracy of the estimated target position.

After Equation \ref{eq:pose estimation}, the reprojection error of camera pose $(\mathbf{r}, \mathbf{t})$ is estimated as follows:
\begin{equation}
e_{\text{proj}}
=
\frac{1}{N} 
\sum_{i=1}^{N}
\Bigl\|\,
\mathbf{x}_i
\;-\;
\widehat{\mathbf{x}}_i
\Bigr\|,
\label{eq:2D reprojection error}
\end{equation}
where $\mathbf{x}_i \in \mathbf{R}^2$ is the $i$-th corner’s detected pixel coordinate, and $\widehat{\mathbf{x}}_i$ is that corner’s projected pixel coordinate given the estimated pose.  

% \paragraph{Distance and Uncertainty.}
% We also define a \textit{distance} from camera to tag as
% \begin{equation}
% \text{distance}
% =
% \sqrt{\,t_x^2
%  + t_y^2
%  + t_z^2},
% \end{equation}
% where $(t_x,t_y,t_z)$ is the translation $\mathbf{t}$ or a related geometric offset in the camera frame.  
The uncertainty $\sigma_{\text{distance}}$ is derived from the re-projection error follows:
\begin{equation}
\begin{aligned}
\sigma_{t_x} &= e_{\text{proj}}\,\frac{t_z}{f_x}, \quad
\sigma_{t_y} &= e_{\text{proj}}\,\frac{t_z}{f_x}, \quad
\sigma_{t_z} &= e_{\text{proj}}\,\frac{t_z}{\sqrt{f_x^2 + f_y^2}}
\end{aligned}
\end{equation}

where $f_x$ is the camera’s focal length in pixels and $t_z$ is the depth in the camera coordinate system. 

The standard deviation for each camera’s distance estimate follows:
\begin{equation}
\sigma_{\text{distance}}
=
\sqrt{
\Bigl(\frac{t_x}{d}\sigma_{t_x}\Bigr)^2
 +
\Bigl(\frac{t_y}{d}\sigma_{t_y}\Bigr)^2
 +
\Bigl(\frac{t_z}{d}\sigma_{t_z}\Bigr)^2
}.
\quad
\end{equation}

When multiple cameras detect the same tag, each camera $j$ provides distance $\mathbf{d}$ and standard deviation $\mathbf{\sigma_{d_j}}$. We combine the distance estimates statistically and consider their variability:
\begin{equation}
d_{\mathrm{fused}}
=
\frac
{
\sum_{j=1}^{m} \frac{d_j}{\sigma_{d_j}^2}
}
{
\sum_{j=1}^{m} \frac{1}{\sigma_{d_j}^2}
}
,\quad
\sigma_{\mathrm{fused}}
=
\sqrt{
\frac{1}{\sum_{j=1}^m \frac{1}{\sigma_{d_j}^2}}
}.
\end{equation}

Consequently, measurements from more reliable cameras (small $\sigma_{d_j}$) dominate the weighted average, and the resulting fused standard deviation $\sigma_{\mathrm{fused}}$ reflects the combined confidence. 

\section{Experimental Setup}
\subsection{Setup}
Three cameras are mounted at complementary angles to provide full 3D coverage of the workspace, with one frontal view and two lateral views to maintain visibility of markers on both the head and TMS coil. Four 24 × 24~mm AprilTag (tag36h11) markers are attached to the head, with two placed near the cheekbones to minimize the effects of facial muscle movement, and additional markers on the forehead and back. A fifth marker is affixed to the center of the TMS coil for tracking of its position and stimulation point.

% At this stage, a dummy head is employed for testing and demonstration purposes, replicating the geometry of a patient's head in a controlled laboratory setting. Using the dummy head lets us evaluate the system’s accuracy, robustness to occlusion, and multi‐camera integration before applying it on real participants, ensuring a reliable setup for TMS experiments. 

We use three commercial cameras (CANYON CNE-CWC5), each with a $1920 \times 1280$ resolution and a 65° wide-angle horizontal field of view, for experiments. Each camera is priced at £21. The implementation details of the experiment are presented in Figure \ref{fig:pipeline}.

\begin{figure}[t]
    \centering
    \subfloat[3D Unity brain model.]{
    \includegraphics[width=0.7\linewidth]{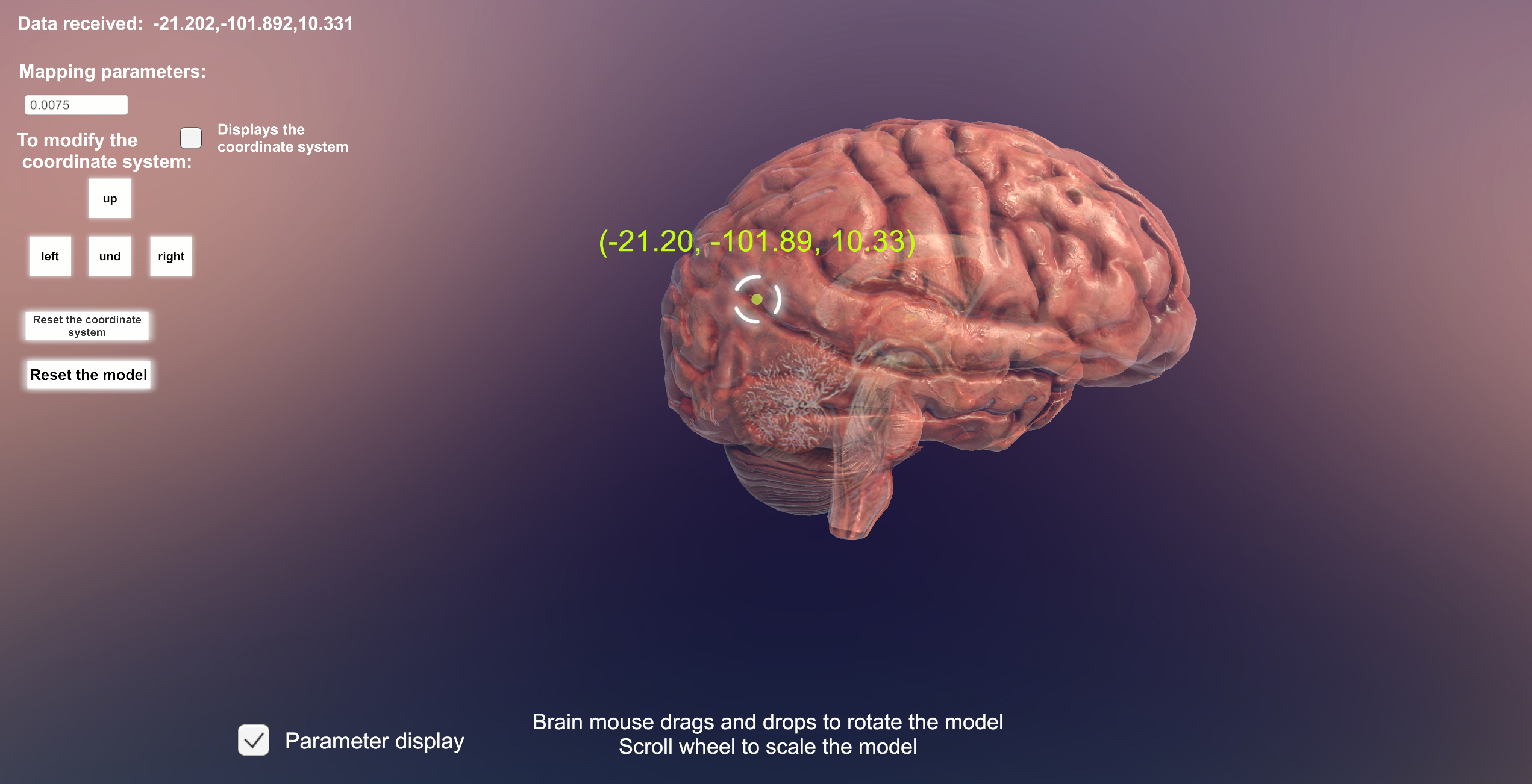}
    \label{fig:3D_brain_model}
    } 
    
    \subfloat[Augmented-reality neuronavigation.]{
    \includegraphics[width=0.7\linewidth]{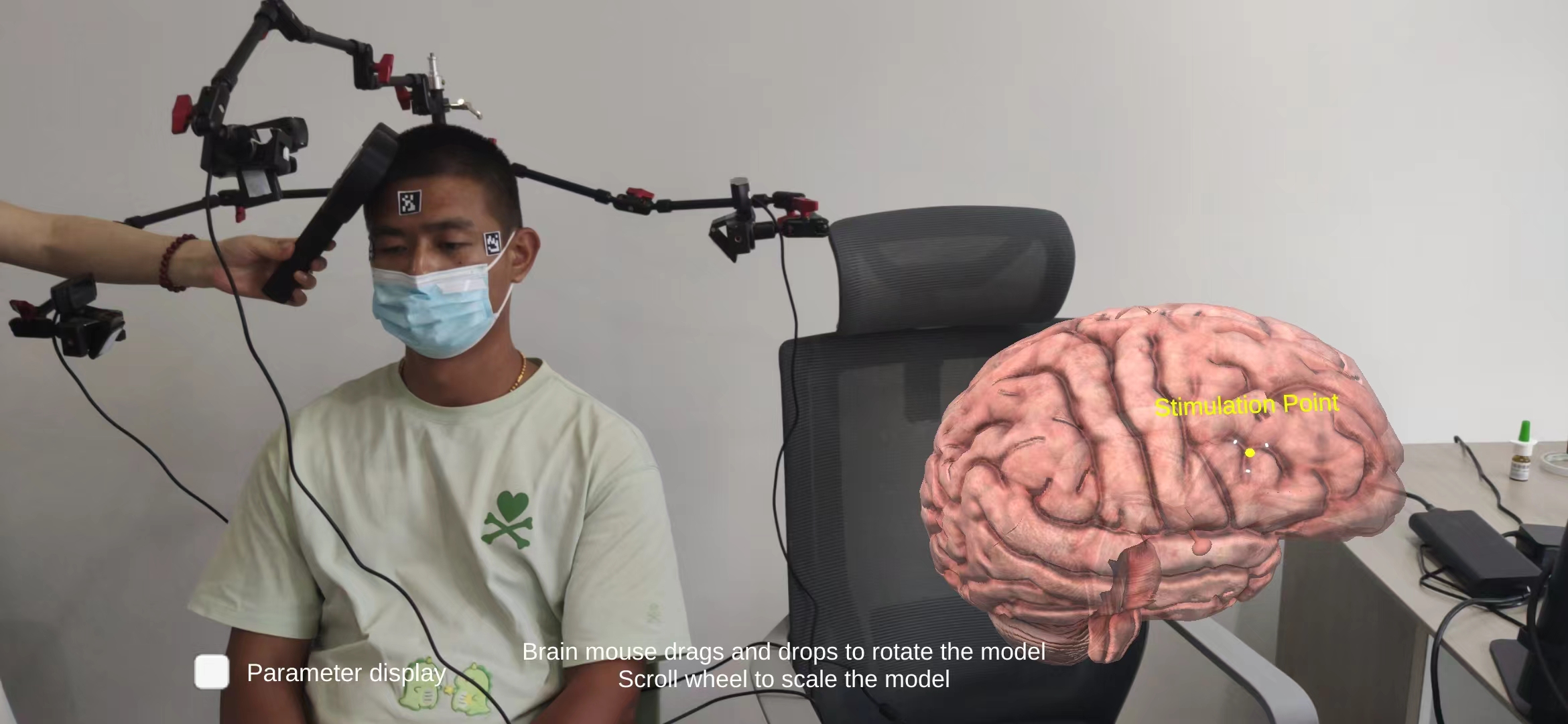}
    \label{fig:AR display}
    } 
    
    \caption{Visualization of the neuronavigation system. (a) Real-time 3D visualization of the TMS stimulation point integrated with a digital brain model in Unity. (b) Augmented reality interface displaying the stimulation point.}
    \label{fig:neuronavigation system display}
    \vspace{-0.3cm}
\end{figure}

\begin{figure*}[tb]
\setlength{\abovecaptionskip}{2pt}
\centering
\includegraphics[width=0.9\linewidth]{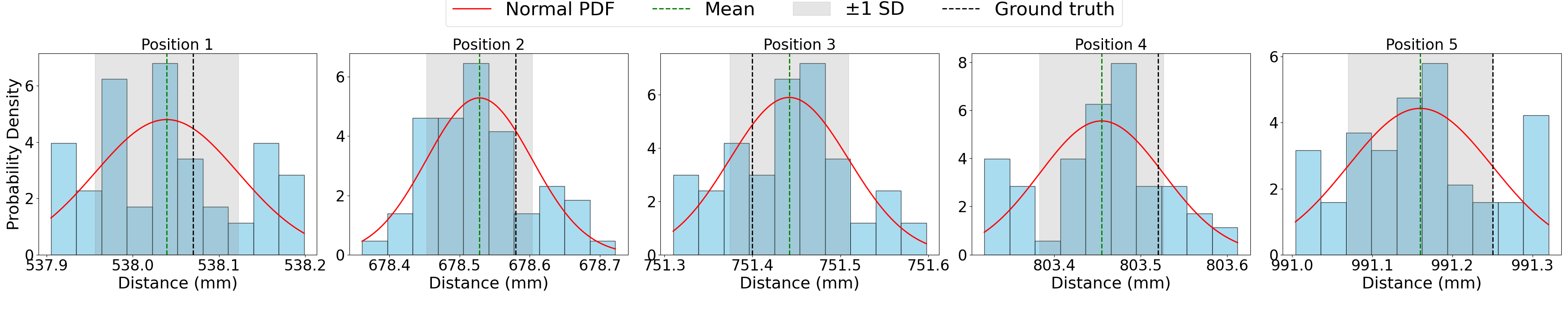}
\caption{\label{fig:distance_R&P}Probability distribution plots of the detected distance in the multi-camera neuronavigation system evaluation across five positions.}
\end{figure*}

\begin{figure*}[tb]
\setlength{\abovecaptionskip}{2pt}
\centering
\includegraphics[width=0.9\linewidth]{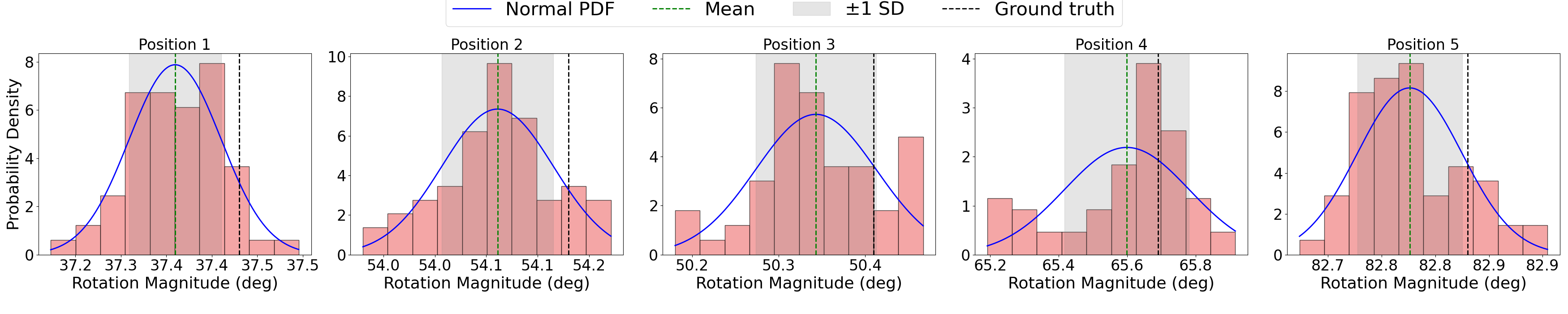}
\caption{\label{fig:angle_R&P}Probability distribution plots of the detected orientation in the multi-camera neuronavigation system evaluation across five positions.}
\end{figure*}

\subsection{Augmented Reality Neuron Navigation System}
To enhance the visualization of the targeted brain region during TMS-guided neuronavigation, we first develop a 3D brain model in Unity that dynamically updates to indicate the current stimulation target based on the real-time optical based multi-camera tracking system using the transmission control protocol‌ (TCP). The estimated simulation point is mapped onto the brain model, allowing clinicians to easily visualize the precise location of stimulation in a virtual environment, as shown in Figure~\ref{fig:3D_brain_model}.

% To further improve usability in clinical settings, we integrate AR technology to project the 3D brain model into the real-world environment, bridging the gap between digital visualization and practical application. By leveraging AR headsets or mobile AR devices, clinicians can interactively view the stimulation target overlaid onto the patient’s head, enhancing spatial precision and ease of operation. Figure \ref{fig:AR display} illustrates the AR Neuron Navigation System. A Xiaomi Mi 9 mobile phone serves as the visualization device. The phone features a 6.39-inch screen with a resolution of 2340 × 1080.

We integrate AR to project a 3D brain model directly into the real-world environment, enabling intuitive visualization of stimulation targets on the patient’s head. AR visualization is implemented using AR Foundation, with tracking data streamed from a PC to an Android device over a local network to ensure real-time responsiveness. Clinicians can thus view and align the TMS coil directly on the patient. The brain model can be either shown alongside the patient or spatially registered and overlaid directly onto the patient’s head, and is scaled to match individual head dimensions. Figure~\ref{fig:AR display} illustrates the AR neuronavigation interface, which is demonstrated using an Android smartphone (Xiaomi Mi~9, $2340 \times 1080$ resolution). By avoiding external monitors or head-mounted AR displays, the system reduces training requirements and supports a more natural, direct view of the stimulation site. This Unity-based AR integration improves usability, accessibility, and precision, making TMS neuronavigation more efficient for clinical practice.

\section{Results} 

\subsection{Precision and Accuracy}
The precision and accuracy experiment evaluates the system’s measurement stability under identical conditions and its absolute error relative to ground truth. Precision reflects variability across repeated measurements, while accuracy indicates closeness to the true position. Measurements were collected by keeping the TMS coil stationary and recording 100 samples at five positions. As shown in Figures~\ref{fig:distance_R&P} and \ref{fig:angle_R&P}, the proposed multi-camera neuronavigation system demonstrates consistently narrow, near-Gaussian distributions, indicating high precision and stable measurement behavior.

Depth estimation remains stable across all tested distances (approximately 530–990~mm), with standard deviations of 0.07–0.09~mm. Absolute depth errors remain below 1~mm, suggesting robust and accurate distance estimation with only minor deviations attributable to residual calibration or tag detection noise. Rotation measurements exhibit similarly strong consistency, with angular standard deviations of 0.04–0.06° and absolute errors well below 1°. The tight Gaussian fits indicate well-controlled angular noise and reliable orientation estimation across positions.

Our system achieves sub-0.1~mm distance precision and sub-0.1° angular precision, with absolute errors below 0.5~mm and 0.3°, respectively. These results confirm the robustness of the multi-camera setup and its ability to deliver high precision and accuracy across diverse measurement conditions.
% Measurements stay within acceptable limits for neuronavigation. The system provides the accuracy and repeatability required for precise tracking in surgical and medical applications.

% \subsection{Position Acquisition Rate}
% In positioning systems for neuronavigation, the position acquisition rate is a critical factor affecting real-time usability and user experience. Given the multi-camera setup used in our system, the computational demand can scale with the number of views, potentially introducing latency. To evaluate latency, we conduct empirical measurements on 50 consecutive frames, recording the system's end-to-end acquisition time for each localization cycle. Figure \ref{fig:position acquisition rate} illustrates real-time position acquisition analysis. The average acquisition time is 0.59~s with a standard deviation of 0.09~s, and 94\% of the trials completed under 0.8~s. The fastest response is 0.34~s, and the slowest is 0.88~s. These results show that the system is generally fast enough for interactive use.

% \begin{figure}[t]
% \centering
% \includegraphics[width=1\linewidth]{pictures/position acquisition rate.png}
% \caption{\label{fig:position acquisition rate} Position acquisition times over 50 trials. The system consistently acquires positions in under one second, with most measurements below 0.8~seconds for interactive applications.}
% \end{figure}

\subsection{Stimulation Point Localization Accuracy}
\begin{figure}[t]
    \centering
    \subfloat[The positioning of 15 designated TMS stimulation targets for testing.]{
    \includegraphics[width=0.75\linewidth]{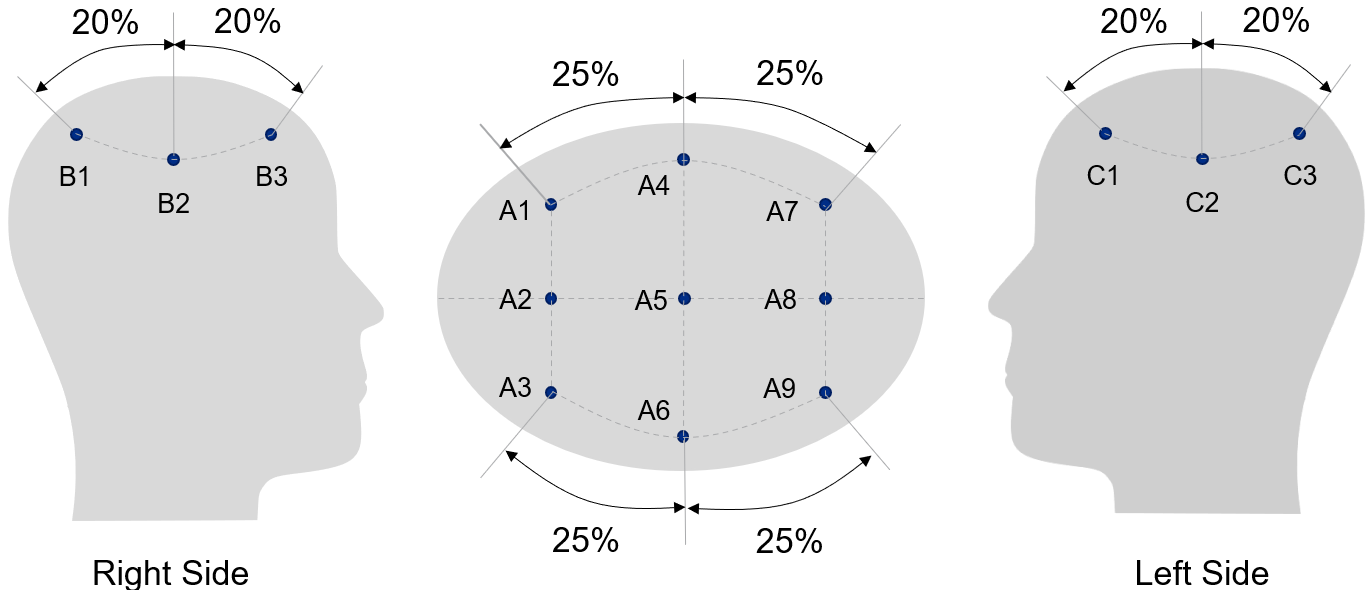}
    \label{fig:accuracy_point_location}
    } 
    
    \subfloat[Bar chart illustrating the stimulation target localization accuracy across 15 different test points. Each bar’s height represents the measured 3D re-projection error in millimeters.]{
    \includegraphics[width=0.75\linewidth]{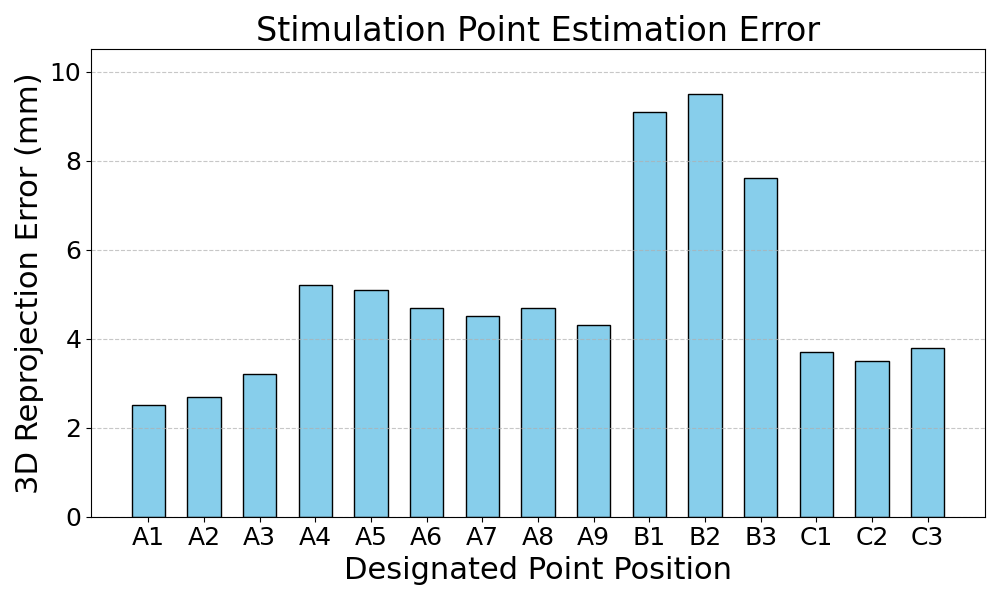}
    \label{fig:accuracy box plot}
    } 
    
    \caption{Selected stimulation target localization accuracy.}
    \label{fig:stimulation point localization accuracy}
    \vspace{-0.3cm}
\end{figure}

We define our system's accuracy through the Euclidean distance between the estimated 3D position, \(\hat{\mathbf{X}}\), and the known ground‐truth position, \(\mathbf{X}_{\text{true}}\). The ground‐truth positions are known locations measured by optical trackers. Figure~\ref{fig:accuracy_point_location} shows the location of 15 selected points for accuracy testing and Figure~\ref{fig:accuracy box plot} illustrates localization accuracy across 15 different test points.

Figure~\ref{fig:accuracy box plot} reveals that the stimulation point estimation error varies across designated positions. This variation in error indicates that there is both systematic bias and random variability in our system. 33.3\% of the measurements have errors below 4\,mm, 33.3\% of the measurements are between 4 and 6\,mm, and 16.7\% exceed 6\,mm. 

\begin{figure}[t]
\centering
\includegraphics[width=0.75\linewidth]{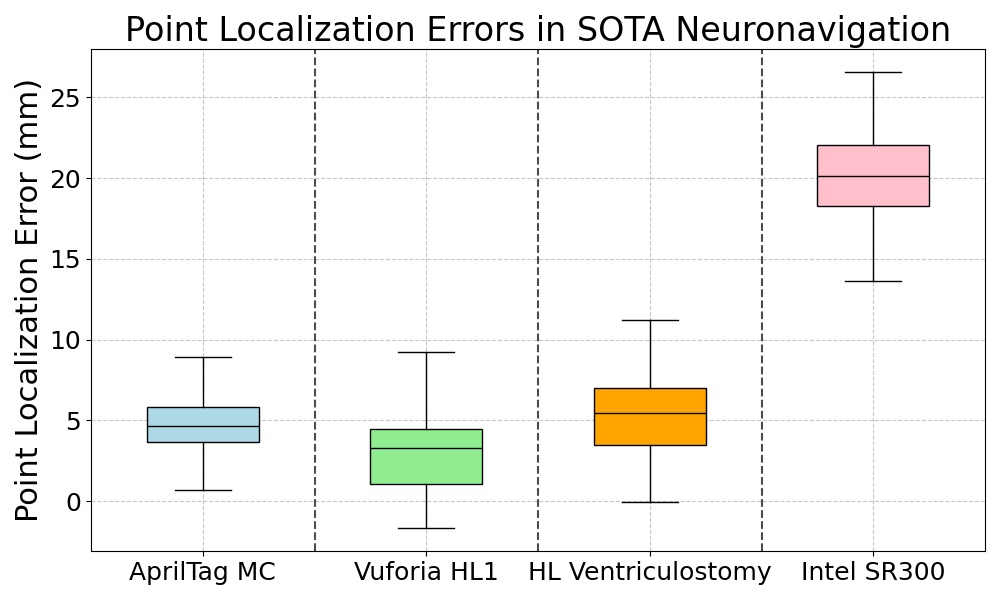}
\caption{\label{fig:accuracy SOTA box plot}Comparison of point localization errors between state-of-the-art and the presented neuronavigation systems.}
\vspace{-0.3cm}
\end{figure}

Figure~\ref{fig:accuracy SOTA box plot} compares the proposed method with three state-of-the-art neuronavigation systems: Vuforia with HoloLens~1~\cite{Frantz2018AugmentingMH}, HoloLens-assisted ventriculostomy~\cite{Schneider2021AugmentedRV}, and Intel RealSense SR300~\cite{asselin2018towards}. These systems represent marker-based AR tracking, head-mounted AR guidance, and depth-sensing-based tracking, respectively. Our approach and Vuforia HL1 achieve the lowest localization errors, with average errors of 4.94~mm and 3.1~mm, respectively. Both rely on marker-based tracking, which provides stable and reliable performance but requires adequate lighting and unobstructed marker visibility. HL-assisted ventriculostomy shows slightly higher error (5.2~mm), likely due to minor spatial distortions introduced by holographic overlays. Intel SR300 exhibits the largest error (20~mm), reflecting the higher sensitivity of depth-based tracking to lighting conditions, surface properties, and viewing angles. Overall, the results indicate that marker-based approaches offer superior localization accuracy compared to depth-sensing methods, supporting the effectiveness of the proposed system for precise neuronavigation.

% Specifically, if we collect repeated measurements \(\hat{\mathbf{X}}^{(1)}, \hat{\mathbf{X}}^{(2)}, \dots, \hat{\mathbf{X}}^{(n)}\), each obtained from pose estimation or reconstruction, then each individual 3D translational error is given by
% \[
% \bigl\lVert \,\hat{\mathbf{X}}^{(i)} - \mathbf{X}_{\text{true}}\bigr\rVert,
% \]
% where \(\lVert \cdot \rVert\) denotes the Euclidean norm. We also combine both the average error and the dispersion into a single Root‐Mean‐Square Error (RMSE), defined as
% \[
% \text{RMSE} \;=\; \sqrt{
%   \frac{1}{n} \;\sum_{i=1}^{n} \bigl\lVert \hat{\mathbf{X}}^{(i)} - \mathbf{X}_{\text{true}} \bigr\rVert^{2}
% }\,.
% \]

% \begin{figure}[h]
% \centering
% \includegraphics[width=0.8\linewidth]{pictures/accuracy point location.png}
% \caption{\label{fig:accuracy point location}The positioning of 15 designated TMS stimulation points for accuracy testing.}
% \end{figure}

% \begin{figure}[h]
% \centering
% \includegraphics[width=0.8\linewidth]{pictures/accuracy box plot.png}
% \caption{\label{fig:accuracy point location}Bar chart illustrating the stimulation point estimation error across 15 different test points (A1–A9, B1–B3, C1–C3). Each bar’s height represents the measured 3D reprojection error in millimeters.}
% \end{figure}

% \section{Use Case in Augmented Reality}

\section{Conclusion and Future Work}
This paper presents an optical tag-based neuronavigation system as a cost-effective and accessible alternative to existing solutions. By using lightweight printed fiducial markers and standard cameras, the system avoids expensive tracking hardware and complex calibration while achieving localization accuracy comparable to state-of-the-art methods. The integration of AR visualization further improves usability and spatial awareness, resulting in a neuronavigation approach that is easier to learn and more efficient. Future work will focus on improving robustness under challenging conditions, including variable lighting, rapid motion, and calibration sensitivity. This may be addressed by refining marker detection, enhancing real-time multi-camera calibration, and integrating hybrid tracking that combines optical tags with depth sensing. In addition, Bayesian filtering could be applied to further reduce measurement noise and improve localization accuracy.

% Compared to depth-sensing systems like Intel RealSense SR300, the AprilTag-based system provides more stable tracking without the inconsistencies caused by lighting conditions or sensor range limitations. The system’s ease of implementation and minimal hardware requirements make it practical for both research and clinical applications. AR visualization enhances user experience by offering intuitive navigation guidance, while neuronavigation hints further improve efficiency by suggesting appropriate stimulation points.

% The authors would like to express their gratitude to Siqi Miao and Kaiyi Chen for their assistance with the drawings.}

%\bibliographystyle{abbrv}
\bibliographystyle{abbrv-doi}

\bibliography{template.bib}
\end{document}